\documentclass[preprint,12pt]{elsarticle}
\usepackage[margin=2.4cm]{geometry}
\usepackage{graphicx}
\usepackage{multirow}
\usepackage{amsmath}
\usepackage{tikz}\usetikzlibrary{decorations.markings}
\begin{document}

\begin{frontmatter}
\title{Linear Approximation of Rear Stagnation-point Flow}
\author{ Chio Chon Kit}
\date{\today}

\begin{abstract}
This paper investigates the nature of the development of two-dimensional steady flow of an incompressible fluid at the rear stagnation-point. 
\end{abstract}

\begin{keyword}
rear stagnation-point flow \sep  similarity solutions \sep  analytical solution 
\end{keyword}

\end{frontmatter}

\section{Introduction}
The classical two-dimensional steady stagnation-point flow on th plane boundary $y=0$ can be analysed exactly by Hiemenz \cite{hiemenz1911grenzschicht}. At a rear stagnation point, 
on a circular cylinder say, the external flow is extracted away from the rear stagnation point. Common observation shows that when the flow is everywhere irrotational. A vortex sheet is formed near the plane and reversed flow develops in the region of vortical flow. 

Forward stagnation point at which a balance is achieved between diffusion of vorcity and the inertia results in a steady solution. On the contrary, due to the advection of diffusion of vorticiy,  rear stagnation-point flows have no steady solution. 

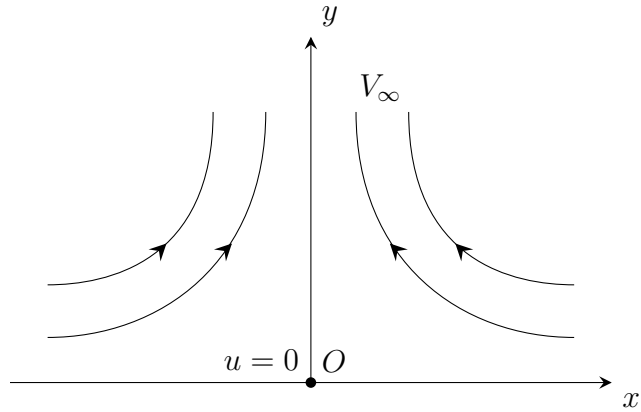
\begin{figure}[!htb]
\centering
\begin{tikzpicture}[>=stealth]
\draw[
    decoration={markings,mark=at position 1 with {\arrow[scale=1.5]{>}}},
    postaction={decorate},
    shorten >=0.4pt
    ] (-4,-0.6) -- (4,-0.6);
\coordinate [label=-45:$x$] (a) at (4,-0.6);
\draw[
    decoration={markings,mark=at position 1 with {\arrow[scale=1.5]{>}}},
    postaction={decorate},
    shorten >=0.4pt
    ](0,-0.6) -- (0,4);
\coordinate [label=45:$y$] (a) at (0,4);
\coordinate [label=45:$V_\infty$] (a) at (0.5,3);
\fill[black] (0,-0.6) circle (2pt);
\coordinate [label=45:$O$] (a) at (0,-0.6);
\node[right=0] at (-1.3,-0.3) {$u=0$};
\draw (0.6,3) .. controls (0.6,1) and (2,0) .. (3.5,0);
\draw[
    decoration={markings,mark=at position 1 with {\arrow[scale=2]{>}}},
    postaction={decorate},
    shorten >=0.4pt
    ] (1.15,1.15) -- (1.05,1.25);
\draw (-0.6,3) .. controls (-0.6,1) and (-2,0) .. (-3.5,0);
\draw[
    decoration={markings,mark=at position 1 with {\arrow[scale=2]{>}}},
    postaction={decorate},
    shorten >=0.4pt
    ] (-1.15,1.15) -- (-1.05,1.25);
\draw (1.3,3) .. controls (1.3,1) and (2.7,0.7) .. (3.5,0.7);
\draw[
    decoration={markings,mark=at position 1 with {\arrow[scale=2]{>}}},
    postaction={decorate},
    shorten >=0.4pt
    ] (2.02,1.15) -- (1.92,1.25);
\draw (-1.3,3) .. controls (-1.3,1) and (-2.7,0.7) .. (-3.5,0.7);
\draw[
    decoration={markings,mark=at position 1 with {\arrow[scale=2]{>}}},
    postaction={decorate},
    shorten >=0.4pt
    ] (-2.02,1.15) -- (-1.92,1.25);
\end{tikzpicture}
\caption{Coordinate system of rear stagnation-point flow}
\label{csp}
\end{figure}
\section{Flow Analysis Model}
We shall demonstrate the linear approximation for the two-dimensional case. We begin with writing the governing equations in conservative velocity form in the Cartesian coordinates:
\begin{subequations}
 \begin{gather}
\frac{\partial u}{\partial x}+\frac{\partial v}{\partial y}=0
\label{eq:e0}
  \end{gather}
 \begin{gather}
    \frac{\partial u}{\partial t}+ u \frac{\partial u}{\partial x}+v\frac{\partial u}{\partial y}=  -\frac{1}{
     \rho}\frac{\partial p}{\partial x}+ \nu \left(\frac{\partial^2 u}{\partial
     x^2}+\frac{\partial ^2u}{\partial y^2}\right)
     \label{e3}\\\notag\\
     \frac{\partial v}{\partial t}+ u \frac{\partial v}{\partial x}+v\frac{\partial v}{\partial y}=  -\frac{1}{
     \rho}\frac{\partial p}{\partial y}+ \nu \left(\frac{\partial^2 v}{\partial
     x^2}+\frac{\partial ^2v}{\partial y^2}\right)
     \label{e4}
  \end{gather}
  \label{e3_4}
\end{subequations}
The equation of continuity (\ref{eq:e0}) is integrated by introducing the stream function $\psi$:
  \begin{equation}
   u=   \displaystyle \frac{\partial \psi}{\partial y}\qquad\mathrm{and}\qquad v= \displaystyle- \frac{\partial \psi}{\partial x}
   \label {stream}
  \end {equation}
In rear stagnation flow without friction (ideal fluid flow), the stream function may be written as
  \begin{equation}
\psi =\psi_\infty = -A_\infty xy
   \label {eq:e00}
\end {equation}
where $A_\infty$ is a constant and from which
  \begin{equation}
U_\infty=-A_\infty x \qquad\mathrm{and}\qquad V_\infty= A_\infty y.
   \label {eq:e01}
 \end {equation}
We have $U_\infty=0$ at $x=0$ and $V_\infty=0$ at $y=0$, but the no-slip boundary at wall $(y=0)$ cannot be satisfied.

In a (real) viscous fluid the flow motion, Proudman and Johnson \cite{proudman1962boundary} model the flows by considering a very simple function of a particular similarity variable
\begin{subequations}
\begin{gather}
\psi = -\sqrt{A\nu}xf(\eta, \tau)\label{psi1}\\
\eta = \sqrt{\frac{A}{\nu}}y\\
\tau = At
\end{gather}
\end{subequations}
Equation (\ref{e3_4}) then gives, for $f(\eta,\tau)$,
   \begin{equation}
    f_{\eta\tau}-(f_{\eta})^2+ff_{\eta\eta}-f_{\eta\eta\eta}=-1,
     \label{e9}
  \end{equation}
with the boundary conditions
  \begin{subequations}
     \begin{gather}
        f(0,\tau)= f_{\eta}(0,\tau)=0\\
        f_{\eta}(\infty,\tau) = 1.~~~~~~~~~~
    \end{gather}
 \label{e10}
  \end{subequations}
At this part it is particular to note that if a steady state is assumed such that $ f_{\eta\tau}\equiv 0$, the resulting equation has no solution.  

\section{Linear Approximation}
Proudman and Johnson have set $V_\infty= 1$ and the corresponding boundary condition $f_{\eta}(\infty,\tau) = 1$. We indicated that for the external flow outside the boundary layer, the hypothesis that the velocity $v(x,\eta,\tau)$ should pass over smoothly into that for inviscid $V_\infty$ is not valid.  The influence of inertia must be taken into account at large distance from the plane. In view of the importance of inertia in the far field, we suggest a linear approximation of the advective terms, which is of dominant importance in the far field, and not in the near field. Consider a substitution
\begin{equation}
u=u' \mathrm{~~~~~and~~~~~}v=V_\infty+v'
\end{equation}
where $(u', v')$ are the Cartesian components of the perturbation velocity, small in the far field. Neglecting the unsteady term, the advective terms become
\begin{equation}
u \frac{\partial u}{\partial x}+v\frac{\partial u}{\partial y}=u' \frac{\partial u'}{\partial x}+v\frac{\partial u'}{\partial y}+V_\infty\frac{\partial u'}{\partial y}
 \label{e11}
\end{equation}
Substituting Equation (\ref{e11}) into Equation (\ref{e3_4}) and neglecting the quadratic terms, we get
\begin{equation}
V_\infty\frac{\partial u'}{\partial y'}=  -\frac{1}{
     \rho}\frac{\partial p}{\partial x}+ \nu \left(\frac{\partial^2 u'}{\partial
     x^2}+\frac{\partial ^2u'}{\partial y^2}\right)
\end{equation}
and a similarity variable
\begin{equation}
\psi = -Axf(y)
\end{equation}
From the definition of the stream function, we have
\begin{equation}
u'=-Axf_y \mathrm{~~~~~and~~~~~}V_\infty+v'=Af
\end{equation}
Equation (\ref{e3_4}) then gives, for $f(y)$,
  \begin{subequations}
 \begin{gather}
      -V_\infty Axf_{yy}=   -\frac{1}{ \rho}
     \frac{\partial p}{\partial x}- A\nu x f_{yyy}    
     \label{e19}\\
  V_\infty Af_{y}=   -\frac{1}{ \rho}
     \frac{\partial p}{\partial y}+ A\nu  f_{yy}
    \label{e20}
  \end{gather}
\end{subequations}
The pressure gradient can be again reduced by a further differentiation equation~(\ref{e20}) with respect to $x$. That is
  \begin{equation}
\frac{\partial^2 p}{\partial x \partial y}=0
  \end{equation}
and equation (\ref{e19}) reduces to
  \begin{equation}
      f_{yyyy}=  \frac{V_\infty}{\nu} f_{yyy}    
  \end{equation}
with the boundary conditions
  \begin{equation}
        f(0)= f_{y}(0)=f_{y}(\infty)=0
 \label{e21}
  \end{equation}
After intergertation,  we obtained an analytical solution of
  \begin{equation}
        f(y) =y+ \frac{\nu}{V_\infty}\left[1-\exp\left({ \frac{V_\infty}{\nu}y}\right)\right]
 \label{e22}
  \end{equation}

Our objective is to obtain a particular solution of the steady rear stagnation-point flow. The solution is obtained in the similarity transformation for steady viscous flows. The first term of (\ref{e22}) shows that the external flow is directed toward the $y-$axis and away from the plane. The appearance of a negative value in the third term in (\ref{e22}) describes a expontential velocity directed toward the wall. 
The function $\displaystyle \exp\left({x}\right)$ has a Taylor series expansion for $x$, that is
  \begin{equation}
\exp(x) = \sum_{n = 0}^{\infty} {x^n \over n!} = 1 + x + {x^2 \over 2!} + {x^3 \over 3!} + {x^4 \over 4!} + \cdots. 
  \end{equation}
Thus, the flow near the boundary becomes
\begin{eqnarray*}
\lim_{y \rightarrow 0} f(y)
& = &y+\frac{\nu}{V_\infty}\left[1-1- \left({ \frac{V_\infty}{\nu}y}\right)-{1 \over 2}\left({ \frac{V_\infty}{\nu}y}\right)^2-\cdots \right] \\
& \simeq & -{1 \over 2}\left({ \frac{V_\infty}{\nu}y}\right)^2 \\
&< & 0
\end{eqnarray*}

The component of velocity normal to the plane is not outward the plane in the region near the rear stagnation point.  The vorticity created at the plane will be convected outward the wall, which spreads the vorticity towards its source at the boundary. An explanation is that an adverse pressure gradient in the region close to the wall leads to a boundary-layer separation and associated flow reversal, and therefore the flow divides into a wall region of reversed flow and an outer region of forward flow. 

The boundary layer thickness, $\delta$, is the distance across a boundary layer from the plane to a point where the flow velocity has essentially reached the free stream velocity, $U_\infty$. This distance is defined normal to the plane, and the point where the flow velocity is essentially that of the free stream is customarily defined as the point where:
\begin{equation}
u(\delta)=0.99U_\infty
 \label{e23}
  \end{equation}
where $U_\infty=-Ax$.
\\\\
From the analytical solution (\ref{e22}), we have
$$
u(\delta)=-Axf_y  =-Ax\left[1-\exp\left({ \frac{V_\infty}{\nu}\delta}\right)\right]
$$
Substituting into equation (\ref{e23}) reduces to
$$
 -0.99Ax =-Ax\left[1-\exp\left({ \frac{V_\infty}{\nu}\delta}\right)\right]
$$
\begin{equation}
 \delta= \frac{\nu}{V_\infty}\ln (0.01)=-\frac{4.61\nu}{V_\infty}
\end{equation}
which implies that $V_\infty$ is a negative value. 

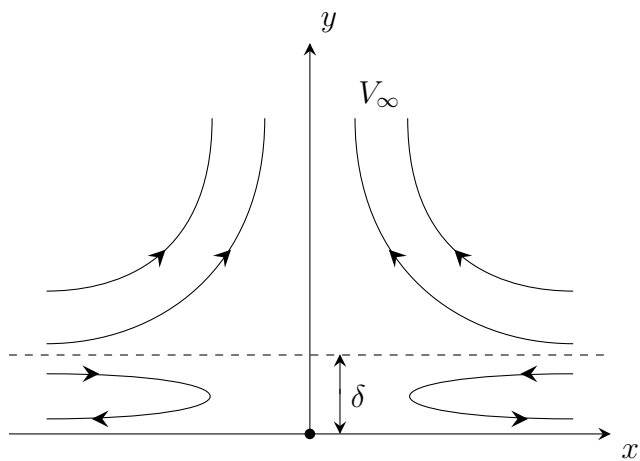
\begin{figure}[!htb]\centering
\begin{tikzpicture}[>=stealth]
\draw[
    decoration={markings,mark=at position 1 with {\arrow[scale=1.5]{>}}},
    postaction={decorate},
    shorten >=0.4pt
    ] (-4,-1.2) -- (4,-1.2);
\coordinate [label=-45:$x$] (a) at (4,-1.2);
\draw[
    decoration={markings,mark=at position 1 with {\arrow[scale=1.5]{>}}},
    postaction={decorate},
    shorten >=0.4pt
    ](0,-1.2) -- (0,4);
\coordinate [label=45:$y$] (a) at (0,4);
\coordinate [label=45:$V_\infty$] (a) at (0.5,3);
\fill[black] (0,-1.2) circle (2pt);

\draw (0.6,3) .. controls (0.6,1) and (2,0) .. (3.5,0);
\draw[
    decoration={markings,mark=at position 1 with {\arrow[scale=2]{>}}},
    postaction={decorate},
    shorten >=0.4pt
    ] (1.15,1.15) -- (1.05,1.25);
\draw (-0.6,3) .. controls (-0.6,1) and (-2,0) .. (-3.5,0);
\draw[
    decoration={markings,mark=at position 1 with {\arrow[scale=2]{>}}},
    postaction={decorate},
    shorten >=0.4pt
    ] (-1.15,1.15) -- (-1.05,1.25);
\draw (1.3,3) .. controls (1.3,1) and (2.7,0.7) .. (3.5,0.7);
\draw[
    decoration={markings,mark=at position 1 with {\arrow[scale=2]{>}}},
    postaction={decorate},
    shorten >=0.4pt
    ] (2.02,1.15) -- (1.92,1.25);
\draw (-1.3,3) .. controls (-1.3,1) and (-2.7,0.7) .. (-3.5,0.7);
\draw[
    decoration={markings,mark=at position 1 with {\arrow[scale=2]{>}}},
    postaction={decorate},
    shorten >=0.4pt
    ] (-2.02,1.15) -- (-1.92,1.25);

\draw (3.5,-0.4) .. controls (0.6,-0.4) and (0.6,-1) .. (3.5,-1);
\draw[
    decoration={markings,mark=at position 1 with {\arrow[scale=2]{>}}},
    postaction={decorate},
    shorten >=0.4pt
    ] (2.9,-0.4) -- (2.8,-0.4);
\draw[
    decoration={markings,mark=at position 1 with {\arrow[scale=2]{>}}},
    postaction={decorate},
    shorten >=0.4pt
    ] (2.8,-1) -- (2.9,-1);
\draw (-3.5,-0.4) .. controls (-0.6,-0.4) and (-0.6,-1) .. (-3.5,-1);
\draw[
    decoration={markings,mark=at position 1 with {\arrow[scale=2]{>}}},
    postaction={decorate},
    shorten >=0.4pt
    ] (-2.9,-0.4) -- (-2.8,-0.4);
\draw[
    decoration={markings,mark=at position 1 with {\arrow[scale=2]{>}}},
    postaction={decorate},
    shorten >=0.4pt
    ] (-2.8,-1) -- (-2.9,-1);
\draw [dashed] (-4,-0.15) -- (4,-0.15);

\draw[
    decoration={markings,mark=at position 1 with {\arrow[scale=1.5]{>}}},
    postaction={decorate},
    shorten >=0.4pt
    ] (0.4,-0.67) -- (0.4,-0.14);

\draw[
    decoration={markings,mark=at position 1 with {\arrow[scale=1.5]{>}}},
    postaction={decorate},
    shorten >=0.4pt
    ]  (0.4,-0.6)--(0.4,-1.2) ;
\coordinate [label=-45:$\delta$] (a) at (0.4,-0.4);
\end{tikzpicture}
\caption{Streamlines of rear stagnation-point flow}
\label{vort}
\end{figure}

At this part it is particular to emphasize a point which seems to been ignored in the analysis. Near the plane or in the boundary layer the phenomenon of reversed flow with boundary-layer separation occurred. Since no information concerning the nature of the flow for finite times has yet been included, there is no justification, theoretical or experimental, for supposing that at large distances from the plane ($\eta \rightarrow \infty$) the velocity $v$ should pass over smoothly into that for inviscid $V_\infty$. Once the reversed flow has occurred, the external boundary condition may be converted to $V_\infty=-A_\infty y$ and our analytical solution (\ref{e22}) becomes
  \begin{equation}
        f(y) =y- \frac{\nu}{V_\infty}\left[1-\exp\left({- \frac{V_\infty}{\nu}y}\right)\right]
 \label{e26}
  \end{equation}
and
\begin{equation}
 \delta= \frac{4.61\nu}{V_\infty}
\end{equation}
The general feature of the predicted streamline pattern is sketched in Figure \ref{vort}.

\section{Conclusion}
For the external flow outside the boundary layer, the hypothesis that the velocity $v(x,\eta,\tau)$ should pass over smoothly into that for inviscid $V_\infty$ is not valid. Separation will occur near the wall as $\eta \rightarrow 0$ and the region of reversed flow will move outward away from the wall. Viscous forces are dominant to decelerate the velocities to zero and ultimately the region of reversed flow does not continue to grow but has finite dimensions. 
\bibliographystyle{ieeetr}	
\bibliography{myrefs}

\begin{thebibliography}{1}

\bibitem{hiemenz1911grenzschicht}
K.~Hiemenz, ``{Die Grenzschicht an einem in den gleichf{\"o}rmigen
  Fl{\"u}ssigkeitsstrom eingetauchten geraden Kreiszylinder, Dingl.
  Polytech},'' {\em J}, vol.~326, pp.~321--410, 1911.

\bibitem{proudman1962boundary}
I.~Proudman and K.~Johnson, ``{Boundary-layer growth near a rear stagnation
  point},'' {\em Journal of Fluid Mechanics}, vol.~12, no.~02, pp.~161--168,
  1962.

\end{thebibliography}

\end{document}